\documentclass[aip,chaos,reprint]{revtex4-1}
\usepackage[usenames,dvipsnames]{color}
\usepackage{graphicx}
\usepackage{amsmath}

\begin{document}
%

\title{Approach to nonequilibrium: from anomalous to Brownian diffusion via non-Gaussianity}
\author{I. G. Marchenko}
\affiliation{NSC \lq\lq Kharkiv Institute of Physics and Technology \rq\rq, Kharkiv 61108, Ukraine}
\affiliation{Institute of Physics, University of Silesia, 41-500 Chorz{\'o}w, Poland}



\author{I. I. Marchenko}
\affiliation{NTU \lq\lq Kharkiv Polytechnic Institute\rq\rq, Kharkiv 61002, Ukraine}

\author{J. {\L}uczka}
\author{J. Spiechowicz}
\affiliation{Institute of Physics, University of Silesia, 41-500 Chorz{\'o}w, Poland}


\begin{abstract}

Recent progress in experimental techniques such as single particle tracking allows to analyze both nonequilibrium properties and \emph{approach to equilibrium}. There are examples showing that processes occurring at finite timescales are distinctly different than their equilibrium counterparts. In this work we analyze a similar problem of \emph{approach to nonequilibrium}. We consider an archetypal model of nonequilibrium system consisting of a Brownian particle dwelling in a spatially periodic potential and driven by an external time-periodic force. We focus on a diffusion process and monitor its development in time. In the presented parameter regime the excess kurtosis measuring the Gaussianity of the particle displacement distribution evolves in a non-monotonic way: first it is negative (platykurtic form), next it becomes positive (leptokurtic form) and then decays to zero (mesokurtic form). Despite the latter fact diffusion  in the long time limit is Brownian, yet non-Gaussian. Moreover, we discover a correlation between non-Gaussianity of the particle displacement distribution and transient anomalous diffusion behavior emerging for finite timescales.

\end{abstract}

\maketitle

\begin{quotation} 

Despite the fact that nonequilibrium systems are widespread at both micro and macro scales their theoretical description is far from being complete. There is no a counterpart of equilibrium statistical physics for nonequilibrium states. Various methods are employed depending on a specific case. One of such approaches is based on the framework of Langevin equation. In the paper we apply this method and study the problem of approach to a final nonequilibrium state for a system consisting of a Brownian particle dwelling in a spatially periodic potential and driven by an external time-periodic force. At non-zero temperature, diffusion of the particle is Brownian (normal) in the long time regime. However, for finite times the transient diffusion process is anomalous and relaxes towards the long-time nonequilibrium state.

\end{quotation}

\section{Introduction} 

Physics of equilibrium systems have already been studied for over a century and a half. Since times of Maxwell, Boltzmann and Gibbs, the founding fathers of equilibrium statistical physics, this theory has been successfully applied for prediction and explanation of a wide variety of phenomena. In the last decades systems that are not in equilibrium have attracted a lot of attention \cite{jag,bianca,prost} mainly due to their relevance for living matter which inherently maintains a nonequilibrium state known as homeostatis \cite{homeo}.

However, theory of nonequilibrium systems is far from being complete in contrast to its equilibrium counterpart. The former is not just a trivial extension of the latter. Despite the fact  that systems out of equilibrium are widespread at both micro and macro scales there is no  single and universal theoretical framework to model them. Moreover, there are no analogues of fundamental principles and relations like an equation of state, a minimum of the free energy, an energy equipartition theorem, a detailed balance symmetry and so on. For equilibrium the system reaches this state independently of the initial conditions \cite{evans, frish}. It is not a  general case for setups far from equilibrium (e.g. if the system is non-ergodic). 

In this work we shall analyze this important issue of approach to nonequilibrium. As an example of such a problem we consider paradigmatic dynamics of a Brownian particle dwelling in a spatially periodic potential and subjected to an external time-periodic force which drives it out of equilibrium \cite{reimann,hanggi2009,denisov2014}. This system approaches in the  long time limit a unique \emph{nonequilibrium} and \emph{nonstationary} state \cite{jung}. The mean square deviation of the particle position as well as its diffusion coefficient will serve as our observable of interest. We will monitor them starting from the initial moment of time up to the long time regime $t \to \infty$.  

We note that recent progress in experimental methods such as single particle tracking \cite{manzo,shen}, e.g. using the stroboscopic techniques, nowadays allows to investigate both nonequilibrium properties and the approach to equilibrium. In particular, in Ref. [\onlinecite{kindermann2017}] the authors performed pioneering experiment and probed diffusion of force-free single atoms in a periodic potential created by an optical lattice. In this system they expectedly observed \emph{Brownian} (normal) diffusion and a \emph{monotonic} convergence  of the displacement distribution to a \emph{Gaussian} one quantified by its excess kurtosis. While for finite times the state of the system is nonequilibrium, for $t\to \infty$ the stationary one is equilibrium. 

In contrast, in this work due to the additional presence of the external time-periodic force the system is no longer equilibrium even in the long time regime. We predict a peculiar evolution of the diffusive behavior while approaching the asymptotic nonequilibrium and nonstationary (i.e. time dependent) state. It is composed of three distinct phases \cite{spiechowicz2016scirep, spiechowicz2017scirep}: initially superdiffusion, then subdiffusion and finally normal diffusion. However, we find that in the  long time asymptotic regime diffusion is Brownian, yet non-Gaussian \cite{granick_pnas,granick_nat,slater,chechkin,bialas2020pre}. Moreover, the excess kurtosis of the corresponding displacement distribution measuring its non-Gaussianity evolves in a non-monotonic manner: starting from negative values towards a positive maximum and then decaying to zero. Most importantly, we reveal the existing correlation between the observed diffusive regime and non-Gaussianity of the position distribution.

The paper is organized as follows. In Sec. II, we formulate the model in terms of the Langevin equation. In Sec. III, we define quantifiers used to characterize the diffusion process. Sec. IV contains a discussion of the results. Sec. V provides summary and conclusions.

\section{Archetypal model}
The model which we are going to study is defined by the Langevin equation \cite{risken,march-chaos,march-pre}
\begin{equation}
	\label{model}
	M\ddot{x} + \Gamma\dot{x} = -U'(x) + F(t) + \sqrt{2\Gamma k_B T}\,\xi(t), 
\end{equation}
where the parameter $\Gamma$ represents the friction coefficient while the dot and the prime denotes differentiation with respect to time $t$ and the spatial coordinate $x$, respectively. It describes a classical Brownian particle of mass $M$ moving in a one-dimensional, spatially periodic potential $U(x)$ and driven by an unbiased and symmetric time-periodic force $F(t)$. The potential $U(x)$ is assumed to be symmetric and possesses a spatial period $L$ and the barrier height $2 \Delta U$, namely, 
\begin{equation}
	\label{potential}
	U(x) = U(x+L) = -\Delta U\cos{\left( \frac{2\pi}{L}x \right)}.
\end{equation}
The external driving force $F(t)$ of amplitude $A$ and angular frequency $\Omega$ is assumed in the simple form,
\begin{equation}
	F(t) = A \sin{(\Omega t + \phi_0)}, 
\end{equation}
where $\phi_0$ is the initial phase value. Thermal equilibrium fluctuations due to interaction of the particle with its environment of temperature $T$ are modeled as $\delta$-correlated Gaussian white noise of vanishing mean value, i.e.
\begin{equation}
	\langle \xi(t) \rangle = 0, \quad \langle \xi(t)\xi(s) \rangle = \delta(t - s), 
\end{equation}
where the bracket $\langle \cdot \rangle$ denotes an ensemble average over thermal noise realizations as well as initial coordinates $x(0)$, velocities $v(0)=\dot{x}(0)$ of the Brownian particle and the initial phase $\phi_0$ of the external driving force. The noise intensity $2\Gamma k_B T$ in Eq. (\ref{model}) follows from the fluctuation-dissipation theorem \cite{kubo1966}, where $k_B$ is the Boltzmann constant.
If $A=0$ the stationary state of the system is thermal equilibrium. If $A \neq 0$, the external force $F(t)$ pumps energy into the system and therefore drives it out of the thermal  equilibrium state. 

Despite its simplicity the model (\ref{model}) describes a rich variety of real physical systems. Historically it has served as the representation of a driven pendulum in viscous environment \cite{risken}, however it also can capture fundamental features of modern setups such as superionic conductors \cite{fulde}, charge density waves \cite{charge}, Josephson junctions and their variations including e.g. the SQUIDS \cite{kautz,spiechowicz2015chaos,spiechowicz2019chaos} as well as cold atoms in optical lattices \cite{lutz2013,denisov2014}, to mention only a few. It has been studied for decades, by many authors, in various contexts, with results published in plenty of journals and books. Is there any reason to still study this system? We note that Eq. (\ref{model}) is one of the simplest models representing nonequilibrium physics which, in contrast to theory of equilibrium systems, is still  far  being  complete.

The dimensionless form of Eq. (\ref{model}) reads
\begin{equation}
	\label{dimless-model}
	\ddot{\hat{x}} + \gamma\dot{\hat{x}} = -\hat{U}'(\hat{x}) + f(\hat{t}) + \sqrt{2\gamma Q} \hat{\xi}(\hat{t}), 
\end{equation}
where
\begin{equation}
	\label{scaling}
	\hat{x} = 2\pi \frac{x}{L}, \quad \hat{t} = \frac{t}{\tau_0}, \quad \tau_0 = \frac{L}{2\pi}\sqrt{\frac{M}{\Delta U}}.
\end{equation}
In this scaling the dimensionless mass $m = 1$ and 
\begin{eqnarray} \label{param}
	\gamma = \frac{\tau_0}{\tau_1}, \quad a = \frac{1}{2\pi}\frac{L}{\Delta U} A,  
\quad \omega = \tau_0 \Omega, \quad Q = \frac{k_B T}{\Delta U}, 
\end{eqnarray}
where $\tau_0$ is the characteristic time related to the period of small oscillations inside the potential well of $U(x)$ and $\tau_1 = M/\Gamma$ is the velocity relaxation time of the free Brownian particle.

The rescaled potential is $\hat{U}(\hat{x}) = U((L/2\pi)\hat{x})/\Delta U = -\cos{\hat x}$. 
The dimensionless external time-periodic force is $f(\hat{t}) = a \cos{(\omega \hat{t} + \phi_0)}$. The rescaled thermal noise reads $\hat{\xi}(\hat{t}) = (L/2\pi \Delta U)\xi(t) = (L/2\pi \Delta U)\xi(\tau_0\hat{t})$ and has the same statistical properties as $\xi(t)$; i.e., $\langle \hat{\xi}(\hat{t}) \rangle = 0$ and $\langle \hat{\xi}(\hat{t})\hat{\xi}(\hat{s}) \rangle = \delta(\hat{t} - \hat{s})$. The dimensionless noise intensity $Q$ is the ratio of thermal energy and half of the activation energy the particle needs to overcome the nonrescaled potential barrier. From now on we shall stick to these dimensionless variables. In order to simplify the notation we omit the hat-notation in Eq. (\ref{dimless-model}).

Let us note that the phase space of the system $\{x, y=\dot x, z = \omega t \}$ is three-dimensional and the parameter space $\{\gamma, a, \omega, Q\}$ is four-dimensional. There are both non-chaotic (regular) and chaotic regimes; locked states in which the motion is confined to a finite number of spatial periods of the potential $U(x)$ and running states when the motion is  unbounded in space. Therefore the system can exhibit extremely rich and unexpected properties in various regions of the four-dimensional parameter space. E.g. in one of such intervals the diffusion coefficient can decrease when temperature increases within a finite temperature window \cite{NJP2016}. The origin of this non-intuitive behaviour is a deterministic mechanism consisting of a few unstable periodic orbits embedded into a chaotic attractor together with thermal noise-induced dynamical changes upon varying temperature. In the regime in which the frequency $\omega$ is less than the natural frequency of oscillations of a particle at the bottom of the potential $U(x)$ well and the relatively small  amplitude of the ac driving  only localized trajectories exist in a deterministic system and there are optimal values of $\omega$ for which the diffusion coefficient can be enhanced by several orders of magnitude \cite{ivan18}. In this paper we consider the case of high frequencies and large amplitudes of the external driving force which is clearly distinct from the previously considered regimes. 

Because neither the Langevin Eq. (\ref{dimless-model}) nor the corresponding Fokker-Planck equation can be solved within analytical means we performed comprehensive numerical simulations of the model. First, the second-order   differential equation (\ref{dimless-model}) has been  converted into a set of two differential equations of first order and next we employed a weak 2nd order predictor-corrector method \cite{platen} implemented on a modern graphics processing unit using CUDA parallel computing platform, for details see \cite{spiechowicz2015cpc}. It allowed us to speed up necessary calculations up to several orders of magnitude as compared to traditional methods. The time step of numerical integration $h = 10^{-2} \times \mathsf{T}$ was scaled by the fundamental period $\mathsf{T} = 2\pi/\omega$ of the dynamics. The quantities of interest were averaged over the ensemble of $2^{20} = 1048576$ system trajectories. Eq. (\ref{dimless-model}) is a second-order stochastic differential equation and therefore we need to specify initial conditions for the particle position $x(0)$ and velocity $v(0)$ as well as phase of the driving force $\phi_0$. Here we applied the simplest choice preserving the symmetry of the system, namely, they are distributed uniformly over the corresponding intervals $x(0) \in [-\pi, \pi]$, $v(0) \in [-2,2]$ and $\phi_0 \in [0,2\pi]$. However, we stress that our findings do not depend on the form of initial conditions since the system of interest is in high temperature regime for which it quickly becomes ergodic and forgets about the initial state \cite{spiechowicz2016scirep}.

\section{Properties of transient diffusion}
Our forthcoming analysis will be focused on the diffusion process of the Brownian particle which is characterized by the mean square deviation (variance) of the particle position $x(t)$, namely,
\begin{equation}
	\label{msd}
	\sigma^2(t) =  \langle \left[x(t) - \langle x(t) \rangle \right]^2 \rangle = \langle x^2(t) \rangle - \langle x(t) \rangle^2.
\end{equation}
For Brownian (normal) diffusion regime $\sigma^2(t)$ is a linear function of time and then the diffusion coefficient can be defined by the relation 
\begin{equation}
	\label{dc}
	D =  \lim_{t \to \infty} \frac{\sigma^2(t)}{2t}.
\end{equation}
For the system described by Eq. (\ref{model}) the asymptotic long time diffusion is \emph{normal} at non-zero temperature $Q \neq 0$. Only this case will be considered below. Although diffusion may not be Brownian for transient regimes nevertheless a time-dependent "diffusion coefficient" $D(t)$ can be defined as 
\begin{equation}
    \label{tdc}
    D(t) =  \frac{\sigma^2(t)}{2t}.
\end{equation}
If the mean square deviation $\sigma^2(t)$ grows in time according to a power-law \cite{klages2008, metzler2014}
\begin{equation}
    \langle \sigma^2(t) \rangle \,\sim  t^{\alpha}
\end{equation}
then the time-dependent diffusion coefficient behaves like
\begin{equation}
	D(t) \sim t^{\alpha - 1}
\end{equation}
and Brownian (normal) diffusion is for $\alpha = 1$. The case $0 < \alpha < 1$ corresponding to subdiffusion means that $D(t)$ is a decreasing function of time. In contrast, for $\alpha > 1$ the coefficient $D(t)$ grows in time indicating the emergence of superdiffusion.

The system described by Eq. (\ref{model}) is dissipative and driven by an external time-periodic force. At non-zero temperature,  in long time limit $t \to \infty$ the system  approaches a  unique \emph{nonequilibrium} and \emph{nonstationary} state being characterized by a temporally periodic marginal probability density for the particle velocity. From the Floquet theory one finds that the latter function assumes the same period $\mathsf{T} = 2\pi/\omega$ as the external driving \cite{jung}. Therefore the average particle velocity can be expressed as
\begin{equation}
	\langle v \rangle = \lim_{t \to \infty} \langle \mathbf{v}(t) \rangle	
\end{equation}
where $\mathbf{v}(t)$ stands for the period averaged velocity
\begin{equation}
	\mathbf{v}(t) = \frac{1}{\mathsf{T}} \int_t^{t+\mathsf{T}} ds \langle v(s) \rangle.
\end{equation}
Since both the periodic potential $U(x)$ and the external driving $a\cos{(\omega t)}$ are symmetric the average particle velocity vanishes 
\begin{equation}
	\langle v \rangle \equiv 0. 
\end{equation}
It does not mean that there are no running trajectories at all. While the latter are permitted, for each of them there exist a corresponding partner transporting the particle in the opposite direction so that both contributions cancel. This constraint implies that the probability distributions $N(x,t)$ and $P(\mathbf{v},t)$ for the position $x(t)$ and period averaged velocity $\mathbf{v}(t)$ of the particle are symmetric if averaged over symmetric distribution of the initial conditions $x(0)$, $v(0)$ and $\phi_0$.

\subsection{Variance of the particle position}
To study the problem of the system approach to its asymptotic long time nonequlibrium state we focus on the weak dissipation regime $\gamma = 0.02$ for which naturally the impact of the external force $a\cos{(\omega t)}$ on the dynamics is the most prominent. Note that the weak damping limit $\gamma \to 0$ via the fluctuation-dissipation theorem typically implies also the weak noise regime $\gamma Q \to 0$. However, the reverse is not necessarily true as the latter can be realized also for vanishing temperature $Q \to 0$ and with the arbitrarily large but fixed friction coefficient $\gamma$.

\begin{figure}[t]
\centering
\includegraphics[width=0.9\linewidth]{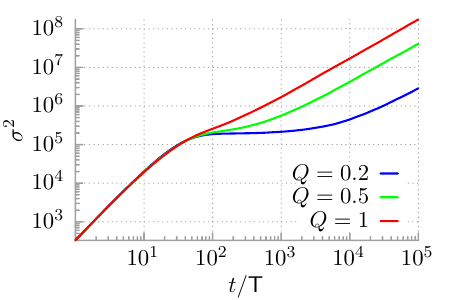}
\includegraphics[width=0.9\linewidth]{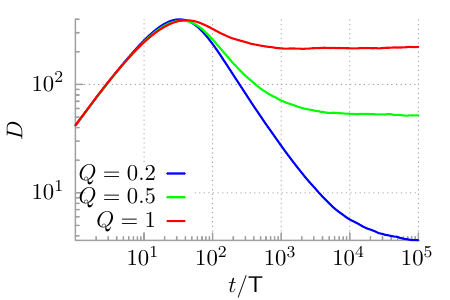}
\caption{Upper panel: The variance  $\sigma^2(t)$ of the particle position $x(t)$. Bottom panel: The time dependent diffusion coefficient $D(t)$. The impact of dimensionless temperature $Q$ is illustrated for $Q=0.2, 0.5, 1$.  Other parameters read: the damping constant $\gamma = 0.02$, the external force amplitude $a = 10$ and its frequency $\omega = 1.59$.}
\label{fig1}
\end{figure}

In Fig. \ref{fig1} we present time evolution of the variance $\sigma^2(t)$ of the particle position $x(t)$ as well as the time dependent diffusion coefficient $D(t)$ for three values of the system temperature $Q = 0.2, 0.5, 1$. At initial stage of the evolution we detect the superdiffusion $\sigma^2(t) \sim t^\alpha$ with $\alpha > 1$. Next, it turns into the subdiffusion $\alpha < 1$ and eventually the Brownian (normal) diffusion is developed with $\alpha = 1$. These three stages of the diffusion process are well illustrated by the time-dependent diffusion coefficient $D(t)$ (bottom panel). Initially $D(t)$ grows with time which corresponds to superdiffusion. Next, it decreases indicating the subdiffusion and finally it tends to the constant value as it is expected for the Brownian diffusion \cite{spiechowicz2016scirep,spiechowicz2017scirep,spiechowicz2019chaos}. 
We want to emphasize that the results presented in Fig. \ref{fig1} are different from those presented in the previous paper \cite{ivan18} in which the authors analyzed the time-dependence of the mean square deviation $\sigma^2(t)$ that exhibits two regimes (superdiffusion and normal diffusion) or three regimes (superdiffusion, next dispersionless regime and finally normal diffusion). In the present paper we study different regime in which superdiffusion, next subdiffusion and finally normal diffusion is detected. However, we stress that the central result of the paper is investigation of the non-Gaussianity of the particle position probability distribution, see below, but not the report of the latter anomalous behavior itself. 

Note that even a small variation of temperature $Q \in [0.2,1]$ significantly affects the subdiffusion lifetime which is very sensitive to this parameter. The duration of anomalous diffusion therefore grows for decreasing temperature \cite{spiechowicz2016scirep}, see the bottom panel of Fig. 1. It is worth to note that for the studied weak dissipation regime $\gamma \ll 1$ the transient anomalous diffusion is long-lived even for the high temperature region $Q \approx 1$. This feature is not expected to appear when thermal noise intensity $\gamma Q$ is much larger. 


\subsection{Probability distributions}

\begin{figure*}[t]
\centering

\includegraphics[width=0.32\linewidth]{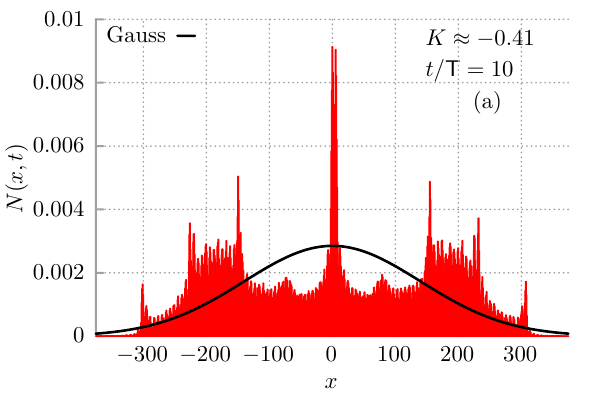}
\includegraphics[width=0.32\linewidth]{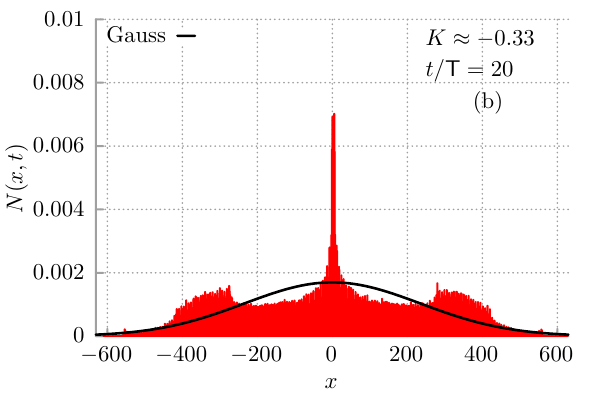}
\includegraphics[width=0.32\linewidth]{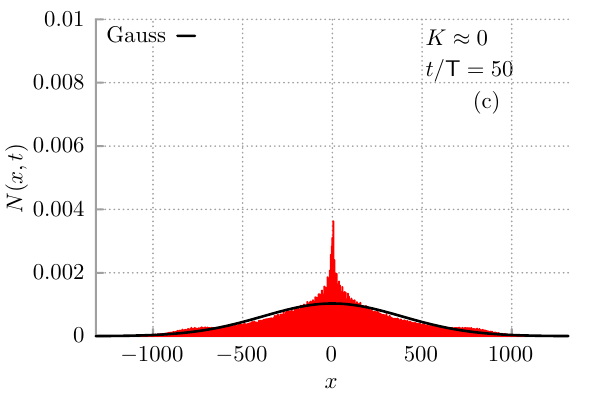}\\

\includegraphics[width=0.32\linewidth]{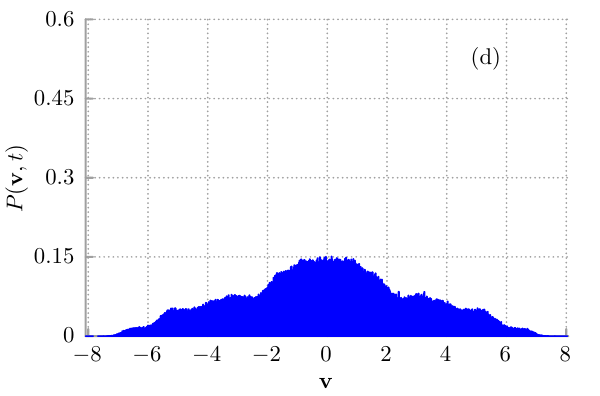}
\includegraphics[width=0.32\linewidth]{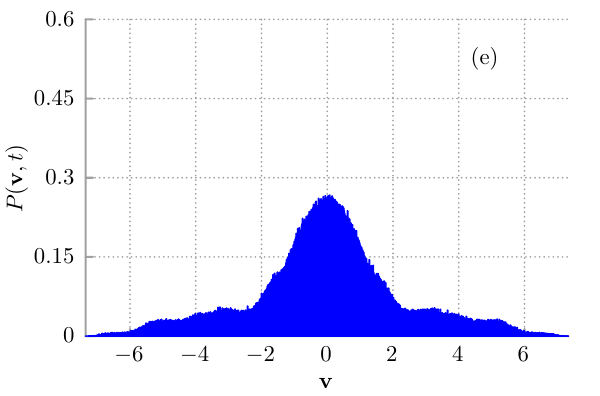}
\includegraphics[width=0.32\linewidth]{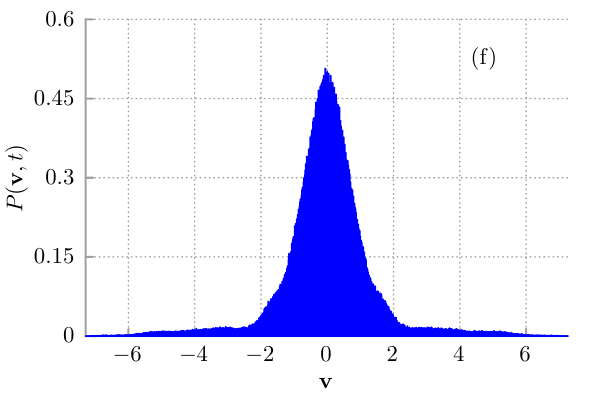}\\

\includegraphics[width=0.32\linewidth]{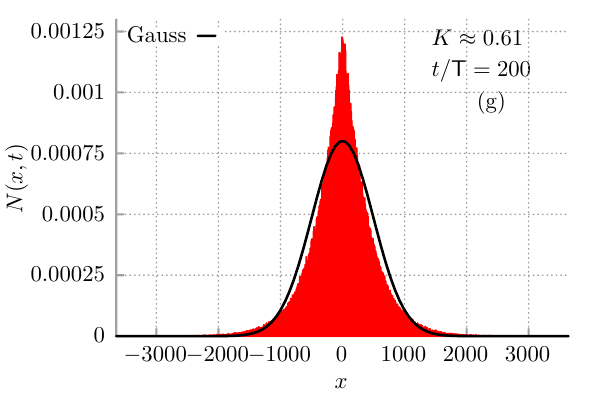}
\includegraphics[width=0.32\linewidth]{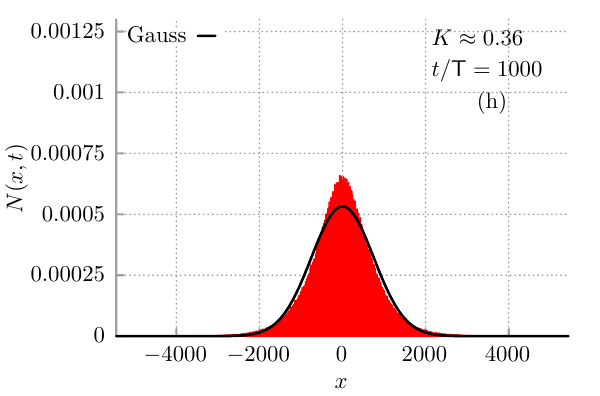}
\includegraphics[width=0.32\linewidth]{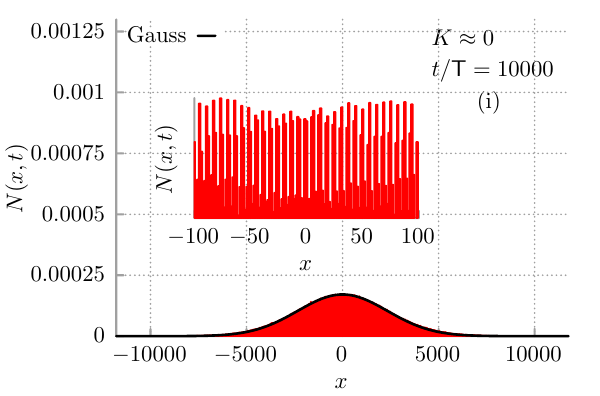}\\

\includegraphics[width=0.32\linewidth]{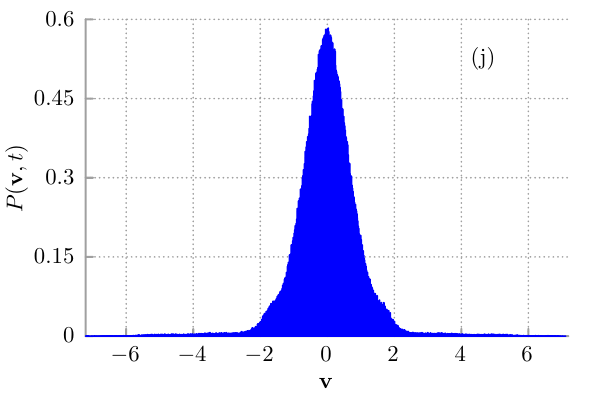}
\includegraphics[width=0.32\linewidth]{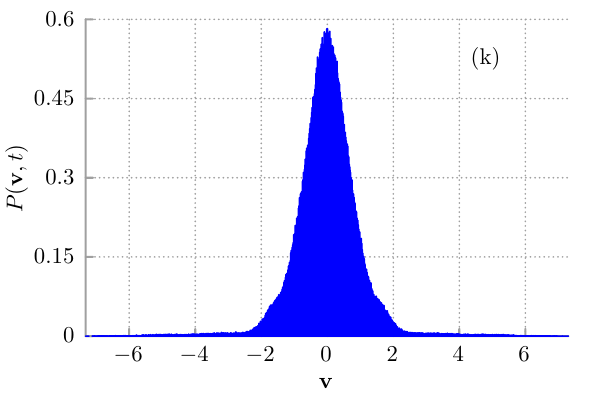}
\includegraphics[width=0.32\linewidth]{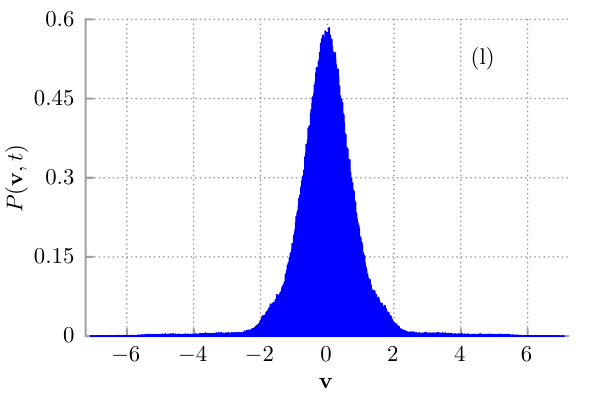}\\

\caption{The probability distributions $N(x, t)$ and $P(\mathbf{v},t)$ for the position $x(t)$ and the period averaged velocity $\mathbf{v}(t)$ of the Brownian particle are shown for growing instants of time. The top two from left to right: $t/\mathsf{T} = 10, 20, 50$, $\mathsf{T} = 2\pi/\omega$. The bottom two from left to right $t/\mathsf{T} = 200, 10^3, 10^4$. The parameters are: $\gamma=0.02, a=10, \omega=1.59, Q = 0.5$. For $N(x,t)$ the corresponding excess kurtosis $K(t)$ is given in the top right corner of the plots. Moreover, the solid black line depicts the fit of the presented distribution to the Gaussian one.}
\label{fig2}
\end{figure*}

Let us now analyze the evolution of the probability distributions $N(x, t)$ and $P(\mathbf{v},t)$ for the position $x(t)$ and the period averaged velocity $\mathbf{v}(t)$ of the Brownian particle at fixed temperature $Q = 0.5$. The corresponding quantities $\sigma^2(t)$ and $D(t)$ characterizing the diffusion process are represented by the green curves in Fig. \ref{fig1}. Time evolution of the probability distributions $N(x, t)$ and $P(\mathbf{v},t)$  is visualised  for selected instances of time in Fig. \ref{fig2}.   We want to detect whether there is any correlation between the properties of $\sigma^2(t)$ or $D(t)$ and features of $N(x, t)$ or $P(\mathbf{v},t)$.

The starting distributions $N(x,0)$ and $P(\mathbf{v},0)$ are rapidly deformed due to the presence of two classes of solutions, namely the locked and running trajectories. The former corresponds to the situation in which motion of the particle is limited to one or several potential wells whereas for the latter it is unbounded. Since the system is driven by thermal noise random transitions between these two classes are possible as well. This structure of solutions is visible in $N(x,t)$, see e.g. panel (a) for $t/\mathsf{T} = 20$. The global maximum at  $x = 0$ corresponds to the locked trajectories oscillating around the potential minimum. The well pronounced local maxima remote from zero $x \neq 0$ are related to the running solutions. On the top of that there are numerous smaller spikes at the minima of the periodic potential $U(x)$ which locally are the most probable particle positions. When time develops first the fine structure of the running solutions is blurred out in $N(x,t)$, see panel (b), and next they begin to disappear in favor of the locked state, see panel (c). These transitions are driven by thermal noise. However, later periods of time reveal another face of thermal fluctuations. They are responsible for diffusive broadening of the distribution $N(x,t)$ and its approach towards the Gaussian-like form, see panels (g)-(i). Strictly speaking it is the Gaussian-like envelope of $N(x,t)$ since numerous spikes at minima of the periodic potential can be still observed and therefore the distribution cannot be Gaussian, see inset in  Fig. 2(i).

This picture is supported by the evolution of the distribution $P(\mathbf{v},t)$ for the period averaged velocity $\mathbf{v}(t)$ of the Brownian particle depicted in Fig. \ref{fig2} as well. For the noiseless case $Q = 0$ corresponding to deterministic counterpart of the studied system we found a set of running trajectories possessing distinct velocities $\langle v \rangle = \pm k \omega$ ($k \in \{1,2,3,4\}$) which strongly break its ergodicity \cite{march-chaos,march-pre}. This set is responsible for the superdiffusion observed at early evolution of $\sigma^2(t)$ or $D(t)$. E.g. if $\langle v \rangle = \pm 4 \omega$ then $x(t) \sim \pm 4 \omega t$ and then diffusion is ballistic $\sigma^2(t) \sim 16 t^2$ with $\alpha = 2$. Of course for non-zero temperature this simple picture is modified by thermal fluctuations that enable transitions between various trajectories which in turn lead to superdiffusion with $1 < \alpha < 2$ instead of the ballistic one $\alpha = 2$. Remnants of the deterministic structure of running trajectories can be easily found in $P(\mathbf{v},t)$ for short times, see panel (d). As time increases these trajectories gradually disappear, see panel (e)-(f), and the population of the locked trajectories grows. It is interesting to note that the characteristic time scale needed for the distribution $P(\mathbf{v},t)$ to reach its asymptotic form is radically shorter than it is the case for $N(x,t)$, c.f. panels (g)-(i) and (j)-(l).

\subsection{Excess kurtosis of the position distribution}

As we demonstrated for the considered parameter regime the diffusion process develops in three stages. Initially for $t \in (0, \mathcal{T}_1)$ the spread of trajectories is  faster than normal and the superdiffusion is observed. Next, in $t \in (\mathcal{T}_1, \mathcal{T}_2)$ diffusion slows down considerably and the subdiffusion is developed \cite{spiechowicz2017scirep}. For $t > \mathcal{T}_2$ the process approaches Brownian, yet non-Gaussian diffusion \cite{granick_pnas,granick_nat}. In the first stage of evolution the probability distribution $N(x,t)$ of the Brownian particle position exhibits several pronounced maxima, see panel (a) of Fig. \ref{fig2}. In the second phase $N(x,t)$ displays only one maximum at the origin $x = 0$, see panel (g) of the same figure. Finally, as time increases further $N(x,t)$ approaches the Gaussian-like form.

To characterize these three phases of the diffusion process we consider the excess kurtosis of the position distribution defined as
\begin{equation}
K(t) = \frac{\langle [x(t)-\mu(t)]^4\rangle}{3 \langle [x(t)-\mu(t)]^2 \rangle^2} -1, \quad \mu(t) = \langle x(t) \rangle =0.
\end{equation}
The excess kurtosis of the Gaussian probability density function is zero. It describes the relation between the distribution outliers, i.e. values that are far away from its mean, to its overall shape. Therefore this quantity measures the degree of "tailedness" in the distribution but \emph{not} its "peakedness". There are examples which are sharply peaked with $K(t) < 0$ as well as those which are flat and yet $K(t) > 0$. However, a distribution with positive excess kurtosis is called leptokurtic and typically possesses fatter tails. In contrast, a case with negative excess kurtosis is called platykurtic and it usually has thinner tails. 

\begin{figure}[t]
\centering
\includegraphics[width=0.9\linewidth]{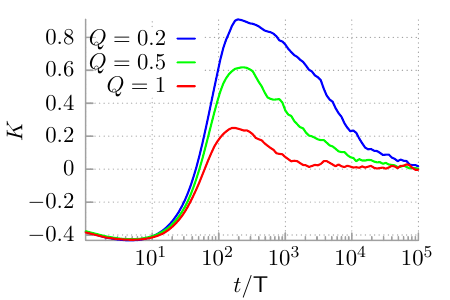}
\includegraphics[width=0.9\linewidth]{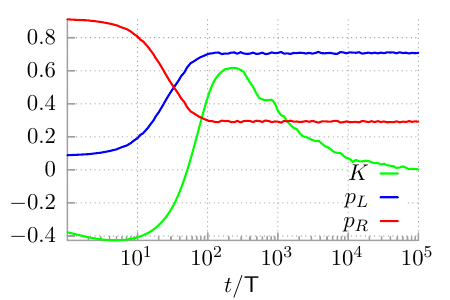}
\includegraphics[width=0.9\linewidth]{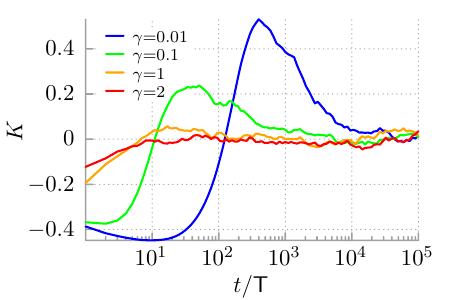}
\caption{Top panel: the time evolution of the excess kurtosis $K(t)$ depicted for different temperature of the system $Q = 0.2, 0.5, 1$. Middle panel: $K(t)$ for $Q = 0.5$ together with the probabilities $p_L(t)$ and $p_R(t)$ of locked and running trajectories, respectively. Bottom panel: $K(t)$ shown for various damping constants $\gamma = 0.01, 0.1, 1, 2$. The remaining parameters are $\gamma=0.02, a=10, \omega=1.59, Q = 0.5$.}
\label{fig3}
\end{figure}

In the top panel of Fig. \ref{fig3} we show time evolution of the excess kurtosis $K(t)$ for different temperature of the system $Q = 0.2, 0.5, 1$, c.f. Fig. \ref{fig1}. Curiously, we can observe there a non-monotonic relaxation towards zero $K(t) = 0$ for which the distribution $N(x,t)$ is mesokurtic. We find that initially the kurtosis is negative $K(t) < 0$ and  decreases up to the time $t \approx 20$ where it attains its minimum. It means that the position outliers are less probable. We note that the scales of the $x$-axes in all panels of Fig. \ref{fig2} are different. It follows that indeed e.g. the probability $\mbox{Pr} \{ x(20) \in (500, 600) \} < \mbox{Pr} \{x(50) \in (500, 600) \}$, c.f. panel (b) and (c) of Fig. \ref{fig2}. Next, the kurtosis starts to increase, for $t \approx 200$ it crosses zero $K(t) = 0$, becomes positive $K(t) > 0$ and grows further up to $t \approx 1000$ where it reaches its maximum. Then the position outliers are much more probable, e.g. $\mbox{Pr} \{ x(200) \in (1000, 1100) \} \gg \mbox{Pr} \{x(20) \in (1000, 1100)\}$, c.f. panel (b) and (g) of Fig. \ref{fig2}. Later, as time goes by, the kurtosis diminishes to zero. 

The interesting observation is that the period in which the kurtosis is negative $K(t) < 0$ corresponds to the interval in which superdiffusion is developed, c.f. Fig.~\ref{fig1}. Similarly, the time window in which the kurtosis is positive $K(t) > 0$ matches the subdiffusive phase of the diffusion process. Finally, in the long time limit $K(t)$ tends to zero and Brownian, yet non-Gaussian diffusion is observed with the Gaussian-like envelope of the distribution $N(x,t)$. However the latter correspondence is not always valid as e.g. for $t = 50\mathsf{T} = 200$ the kurtosis $K(200) \approx 0$ but neither $N(x,t)$ resembles the Gaussian form nor the diffusion is Brownian, see Fig. \ref{fig1} and panel (c) of Fig. \ref{fig2}.

\section{Discussion}

Let us now make several remarks about the presented results. Firstly, the time dependence of the excess kurtosis $K(t)$ can be divided into two intervals, see Fig. \ref{fig3},  (i) $[0,t_K]$ in which this quantity approaches its maximum value $K(t_K)$; (ii) $[t_K,\infty]$ when kurtosis tends to zero. In the middle panel of Fig. \ref{fig3} we show that the first interval $[0,t_K]$ is related to transitions between the locked and running states as a result of which the numbers of these  trajectories are established. Indeed, the probabilities $p_L(t)$ and $p_R(t)$ of the particle to reside in the locked and running state presented there initially rapidly change and saturate to a constant value around $K(t_K)$. In this phase the distribution $N(x,t)$ for the particle position undergoes the most prominent transformation due to the above process.
In the second interval $[t_K,\infty]$ the number of locked and running trajectories is no longer modified and one observes typical spreading of the distribution $N(x,t)$ characteristic for thermal fluctuations influence.

Secondly, in the top panel of Fig. \ref{fig3} the impact of temperature $Q$ on the relaxation of the excess kurtosis $K(t)$ is presented. The second interval $[t_K,\infty]$ when $K(t)$ tends to zero is particularly sensitive to temperature variation. However, $[0,t_K]$ is also affected. In particular, time needed for the distribution $N(x,t)$ to approach its asymptotic Gaussian-like form tends to infinity in the limit of vanishing temperature $Q \to 0$.

Thirdly, in the bottom panel of Fig. \ref{fig3} we present the influence of the damping coefficient $\gamma$ on the evolution of the excess kurtosis $K(t)$ for temperature $Q = 0.5$. One immediately finds that the non-monotonic relaxation of $K(t)$ is detected only for the weak dissipation regime $\gamma \to 0$. It ceases to exist already for $\gamma \approx 1$. It is expected since the influence of the external time-periodic force $a\cos{(\omega t)}$ which drives the system out of thermal equilibrium is most pronounced for the weak damping and in such a case the lifetime of running trajectories is longer. Moreover, the time interval during which the excess kurtosis is negative $K(t) < 0$ is extended when the dissipation constant $\gamma$ decreases. On the other hand, regardless of the friction coefficient value for any finite time (also in the  long time regime) one observes Brownian, yet non-Gaussian diffusion because $N(x, t)$ possesses infinitely many extrema corresponding to the maxima and minima of the spatially periodic potential $U(x)$. If time increases $N(x,t)$ spreads out, i.e. the global maximum at $x = 0$ tends to zero and so do the other extrema while at the same time $N(x,t)$ becomes wider, and finally for  $t \to \infty$ the probability distribution $N(x,t) \to  0$. 


Fourthly, we note that the \emph{non-monotonic} evolution of the excess kurtosis $K(t)$ with the global negative minimum and positive maximum should be contrasted with the \emph{monotonic} decay of $K(t)$ predicted theoretically and observed experimentally for the equilibrium counterpart of the studied system, i.e. without the external driving force $a = 0$ \cite{kindermann2017}. Moreover, in the latter case the kurtosis is only positive $K(t) > 0$. For this reason we conclude that the negative $K(t)$ is characteristic for nonequilibrium states.

\section{Conclusions}

In this work we analyze the problem how a system consisting of a Brownian particle dwelling in a spatially periodic potential and driven by an external time periodic force relaxes to its asymptotic long time nonequilibrium state. In doing so we focused only a specific aspect of this issue, namely the process of diffusion. 

It is known that for this nonequilibrium system diffusion in the long time regime is normal (Brownian). However, by investigating the mean square deviation of the particle position we revealed that when approaching the asymptotic nonequilibrium state diffusion goes through transient but long-lived periods of superdiffusion and subdiffusion.

We described this diffusion process by investigating the probability distribution for the particle position and the period averaged velocity as well as the excess kurtosis of the coordinate density. For the weak dissipation regime  we discovered a characteristic non-monotonic time dependence of the excess kurtosis and revealed its correlation with the diffusive properties of the system. 

It turned out that when superdiffusion is developed the kurtosis is negative, next for subdiffusion it is positive, reaches its maximum and then monotonically decreases to zero when the normal diffusion is observed. Although the kurtosis vanishes the probability distribution for the particle position is only Gaussian-like and therefore asymptotically diffusion is Brownian, yet non-Gaussian. 

In conclusion, from the above findings we reveal how diffusion can evolve from anomalous to Brownian one via non-Gaussianity of the particle displacement probability distribution and the way in which the system approach its asymptotic nonequlibrium state.
   
\section*{Acknowledgment}
This work was supported by the Grants NCN 2022/45/B/ST3/02619 (J.S.). I. G. M. acknowledges University of Silesia for its hospitality since the beginning of the war, 24 February 2022.

\section*{Conflict of interest}
The authors have no conflicts to disclose.

\section*{Data availability statement}
The data that support the findings of this study are available from the corresponding author upon reasonable request.

\end{document}